\documentclass[12pt]{article}

\makeatletter
\renewcommand{\@seccntformat}[1]{{\csname the#1\endcsname.\quad}}
\makeatother
\usepackage{theorem}

\newtheorem{theorem}{Theorem}[section]
\newtheorem{lemma}[theorem]{Lemma}
\newtheorem{corollary}[theorem]{Corollary}
\newtheorem{proposition}[theorem]{Proposition}

\newtheorem{claim}[theorem]{Claim}

{\theorembodyfont{\rmfamily}
\newtheorem{definition}[theorem]{Definition}
\newtheorem{remark}{Remark}

}

\newcommand{\beginproof}{\medskip\noindent{\bf Proof.~}}
\newcommand{\beginproofof}[1]{\medskip\noindent{\bf Proof of #1.~}}
\newcommand{\beginproofdotless}{\medskip\noindent{\bf Proof}}
\newcommand{\finishproof}{\hspace{0.2ex}\rule{1ex}{1ex}}
\newenvironment{proof}{\beginproof}{\unskip\nolinebreak\finishproof\par\medskip}
\newenvironment{proofof}[1]{\beginproofof{#1}}{\unskip\nolinebreak
\finishproof\par\medskip}

\newcounter{tenumerate}

\newlength{\mathindent}
\newcommand{\refeq}[1]{{\rm (\ref{#1})}}
\newlength{\longeqmarginwidth}
\newlength{\longeqwidth}
\newlength{\longeqskiplength}
\def\gr
{\mathrel{\kern0.1em\hbox{\vbox{\baselineskip =-100pt\lineskip=.2ex%
\halign{##\cr\vbox{\hrule height .35pt width 1em} \cr \kern .33em \hbox{$
\scriptscriptstyle  \circ$} \cr \vbox{\hrule height .35pt width 1em}
\cr}}\kern-0.22em}}}

\newcommand{\Vzero}[2]{\mathord{\stackrel{\textstyle\kern1pt\circ}
{\smash V\vbox to6pt{}}}\vphantom{V}_{#1}^{#2}}

\newcommand{\function}[2]{:#1 \longrightarrow #2}
\newcommand{\of}[1]{\left( #1 \right)}

\newcommand{\set}[2]{\left\{\hspace{0.2ex} #1 \left|\: #2
\right. \right\}}

\newcommand{\df}{\stackrel{\rm def}{=}}

\newcommand{\Tr}{{\rm Tr}}

\newcommand{\scr}{\mathcal}

\newcounter{operator}

\newcommand{\reals}{\mathbb{R}}

\usepackage{amsfonts}
\usepackage{amsmath}

\newcommand{\Input}{\mbox{Input}}
\newcommand{\Output}{\mbox{Output}}
\newcommand{\Trace}{\mbox{Trace}}
\renewcommand{\Tr}{\mbox{Tr}}
\newcommand{\Conv}{\mbox{Conv}}
\newcommand{\fnorm}[1]{||#1||_{\rm F}}
\newcommand{\trnorm}[1]{||#1||_{\rm tr}}

\title{Quantum Communication Complexity of Symmetric Predicates}
\author{Alexander A. Razborov \thanks{Institute for Advanced Study,
Princeton, US, on leave from Steklov Mathematical Institute, Moscow,
Russia, razborov@ias.edu. Supported by The von Neumann Fund.}}
\begin{document}
\maketitle

\begin{abstract}
We completely (that is, up to a logarithmic factor) characterize the
bounded-error quantum communication complexity of {\em every} predicate
$f(x,y)$ depending only on $|x\cap y|$ ($x,y\subseteq [n]$). Namely, for a
predicate $D$ on $\{0,1,\ldots,n\}$ let $\ell_0(D)\df
\max\set{\ell}{1\leq\ell\leq n/2\land D(\ell)\not\equiv D(\ell-1)}$ and
$\ell_1(D)\df \max\set{n-\ell}{n/2\leq\ell < n\land D(\ell)\not\equiv
D(\ell+1)}$. Then the bounded-error quantum communication complexity of
$f_D(x,y) = D(|x\cap y|)$ is equal (again, up to a logarithmic factor) to
$\sqrt{n\ell_0(D)}+\ell_1(D)$. In particular, the complexity of the set
disjointness predicate is $\Omega(\sqrt n)$. This result holds both in the
model with prior entanglement and without it.
\end{abstract}

\section{Introduction}

The model of communication complexity, originally introduced by Yao
\cite{Yao2} has since evolved into a very intriguing and important branch
of computational complexity that in particular links and unifies many
different things. In this model, Alice holds an input $x\in X$, Bob holds
$y\in Y$, and they exchange messages to evaluate a Boolean predicate
$f\function{X\times Y}{\{0,1\}}$. The complexity is measured by the number
of bits exchanged, and, like in many other areas of computational
complexity, one distinguishes between deterministic and probabilistic
modes.

Just like the circuit complexity is quite often concerned with symmetric
Boolean functions one class of problems that attracted a considerable
interest in communication complexity is made by {\em symmetric predicates}
which we define as those for which $x,y$ are finite sets and $f_D(x,y) =
D(|x\cap y|)$ for some predicate $D$ on integers. Two most prominent
members of this class are the disjointness predicate $DISJ_n$ ($D(s)\equiv
(s=0)$) and the inner product function $IP_n$ ($D(s)\equiv s\pmod 2$). The
rank lower bound by Mehlhorn and Schmidt \cite{MS} immediately implies a
tight $\Omega(n)$ lower bound on the deterministic communication
complexity of both $DISJ_n$ and $IP_n$.

For the randomized algorithms, \cite{Vaz,ChG,BFS} proved an $\Omega(n)$
lower bound on the complexity of the inner product $IP_n$, and \cite{BFS}
also contained an $\Omega(\sqrt n)$ lower bound for $DISJ_n$. The latter
bound was improved to the optimal $\Omega(n)$ in \cite{KaS}, and their
proof was further simplified in \cite{dis}.

\medskip
The model of {\em quantum} communication complexity was also
introduced by Yao \cite{Yao5}. Suppose that Alice and Bob can employ the
laws of quantum mechanics and are allowed to exchange qubits instead of
classical bits. Can it help them to reduce the amount of communication?

Buhrman, Cleve and Wigderson \cite{BCW} observed that the rank lower bound
for deterministic protocols extends to the quantum case (so, after all the
answer for such protocols can be ``NO''). In particular, both $DISJ_n$ and
$IP_n$ require $\Omega(n)$ qubits to be exchanged by quantum deterministic
(= zero-error) protocols. The rank lower bound was extended in \cite{BuW}
to the stronger model with prior entanglement previously introduced in
\cite{ClB} (in that model, Alice and Bob share an unlimited number of
entangled EPR-pairs before the communication even begins).

\smallskip
The question about the complexity of protocols that allow a small error is
by far more interesting. As far as lower bounds are concerned, Kremer
\cite{Kre}, based upon some ideas from the seminal paper \cite{Yao5},
proved an $\Omega(n)$ lower bound for $IP_n$. This result was extended to
the model with prior entanglement in \cite{CVN*}. Klauck \cite{Kla} looked
at the threshold predicates ($D(s)\equiv (s\geq\ell)$) and exact-$\ell$
predicates ($D(s)\equiv (s=\ell)$) and proved an $\Omega(\ell/\log\ell)$
bound in both cases (without entanglement). The only {\em general} lower
bound for $DISJ_n$ (that corresponds to $\ell=0$) prior to this work was
$\Omega(\log n)$ \cite{AST*,BuW}; we can also mention some partial results
in this direction such as bounds for constant-round protocols \cite{KNT*},
protocols with exponentially small error \cite{BuW} and some highly
structured protocols \cite{HoWo}.

On the upper bounds frontier, the elegant paper \cite{BCW} established a
strong connection between quantum search and quantum communication by
showing how to convert {\em every} quantum search algorithm for {\em any}
Boolean function $g$ into a quantum communication algorithm for the
associated predicate $f_g(x,y)=g(x\cap y)$ with only a logarithmic delay.
Plugging into this procedure Grover's search algorithm \cite{Gro}
immediately gave an $O(\sqrt n\log n)$ upper bound on the bounded-error
quantum communication complexity of disjointness (that was later slightly
improved in \cite{HoWo} to $O(\sqrt n\exp(\log^\ast n))$). \cite{BBC*}
proved that the quantum query complexity of every symmetric Boolean
function $g$ is equal, up to a constant factor, to its {\em approximate
degree} $\widetilde{\deg}(g)$ (defined as the minimal degree of a real
polynomial approximating $g$ on $\{0,1\}^n$ in the $\ell_\infty$-norm
within accuracy $1/3$). Combined with the BCW-reduction, this implies an
$O(\widetilde{\deg}(g)\log n)$ upper bound on the bounded error quantum
communication complexity of $f_g(x,y)$.

\bigskip
In this paper we prove that for {\em every} symmetric predicate $f_D(x,y)$
this communication algorithm is essentially optimal provided we take care
of one ``degenerate'' case. More specifically, let
\begin{equation} \label{l0}
\ell_0(D)\df \max\set{\ell}{1\leq\ell\leq n/2\land D(\ell)\not\equiv
D(\ell-1)}
\end{equation}
and
\begin{equation} \label{l1}
\ell_1(D)\df \max\set{n-\ell}{n/2\leq\ell < n\land D(\ell)\not\equiv
D(\ell+1)}.
\end{equation}
If we let $g_D(x_1,\ldots,x_n)=D(|x|)$, then the classical result by
Paturi \cite{Pat} says that
$\widetilde{\deg}(g_D)=\theta(\sqrt{n(\ell_0(D)+\ell_1(D))})$ which
implies, via \cite{BCW}, an upper bound of $O((\sqrt{n\ell_0(D)}+
\sqrt{n\ell_1(D)})\log n)$ on the quantum bounded-error communication
complexity of $f_D$. This can be easily improved to $O((\sqrt{n\ell_0(D)}+
\ell_1(D))\log n)$ (large values of $|x\cap y|$ are taken care of by the
trivial algorithm in which Alice sends to Bob her entire input). We prove
the lower bound $\Omega(\sqrt{n\ell_0(D)}+ \ell_1(D))$ matching this upper
bound up to a logarithmic factor (Theorem \ref{main_theorem}). Our lower
bound works also in the model with prior entanglement.

\medskip
For the proof of our result we use a multi-dimensional version of the
ordinary discrepancy method (Section \ref{discrepancy:sec}). That is, we
measure the communication matrix against several probability distributions
at the same time. This allows us to reduce our problem to a classical
problem in the discrete polynomial approximation that, quite fortunately,
was solved in the above-mentioned paper \cite{Pat} (Section
\ref{final:sec}). Another specific feature of our approach is that we tend
to apply spectral methods (as opposed to combinatorial ones) more
systematically than it was done in the previous papers on the subject
(this becomes especially critical for handling prior entanglement). In
particular, we show a general lower bound on the quantum communication
complexity of a function in terms of the approximate trace norm of its
communication matrix (Section \ref{tracenorm:sec}).

In the rest of the paper we formulate and prove our main result. Whenever
possible, we try to present in reasonable generality those intermediate
steps in our proof that might be of independent interest.

\section{Quantum communication model and the main result}
\label{main}

There are several equivalent definitions of the quantum communication
model; in our description we follow \cite{BuW} as this variant seems to be
the most convenient to work with.

Let $X,Y$ be finite sets and $f\function{X\times Y}{\{0,1\}}$ be a Boolean
predicate. Let $\scr H_A,\scr C,\scr H_B$ be finite-dimensional Hilbert
spaces representing Alice's part, the channel and Bob's part,
respectively. Like in \cite{BuW} we require that $\scr C$ consists of a
single qubit (that is, $\dim(\scr C)=2$, and $|0\rangle ,\ |1\rangle $ is
its orthonormal basis).

The models with or without prior entanglement differ only in the
unitary vector $\Input(x,y)\in \scr H_A\otimes\scr C\otimes\scr H_B$
prepared at the beginning of the communication. We postpone its definition
and describe first how the communication proceeds. A {\em $c$-qubit
communication protocol} is completely determined by unitary operators
$U_1,U_2,\ldots,U_c$, where $U_i$ acts on $\scr H_A\otimes\scr C$ if $i$
is odd, and on $\scr C\otimes\scr H_B$ if it is even. The output (unitary)
vector is then
\begin{equation} \label{up}
\Output(x,y)\df\ldots (U_3\otimes I_B)(I_A\otimes U_2)(U_1\otimes
I_B)\Input(x,y),
\end{equation}
where $I_A,I_B$ are identity operators on $\scr H_A,\scr H_B$,
respectively. The {\em acceptance probability} of this protocol on $x,y$
is the result of the measurement of $\Output(x,y)$ with respect to $\scr
C$, i.e., the squared $\ell_2$-norm of its orthogonal projection onto
$\scr H_A\otimes |1\rangle \otimes \scr H_B$.

We are still left to describe $\Input(x,y)$. In the model without prior
entanglement, $\scr H_A$ has the orthonormal basis $\set{|a,x\rangle
}{a\in W_A,\ x\in X}$, where $W_A$ is a finite set with a distinguished
element 0 (representing Alice's internal computations). Likewise, $\scr
H_B$ has the orthonormal basis $\set{|y,b\rangle }{b\in W_B,\ y\in Y}$ and
$\Input(x,y)\df |0,x\rangle |0\rangle |y,0\rangle $.

\begin{sloppypar}
In the model with prior entanglement, $\scr H_A$ has the basis
$\set{|a,x,e\rangle }{a\in W_A,\ x\in X,\ e\in E}$ and $\scr H_B$ has the
basis $\set{|e,y,b\rangle }{b\in W_B,\ y\in Y,\ e\in E}$, where $E$ is a
new finite set (corresponding to all possible pure states of entangled
EPR-pairs). The beginning state in this case is
\begin{equation} \label{input_string}
\Input(x,y)\df \frac 1{|E|^{1/2}}\sum_{e\in E}|0,x,e\rangle |0\rangle |e,y,0\rangle .
\end{equation}
It is important that in this model we do not have any control of $|E|$
whatsoever, and it must not appear in our bounds.
\end{sloppypar}

\smallskip
A quantum protocol {\em computes $f(x,y)$ with error $\epsilon$} if its
acceptance probability on every $(x,y)$ is at most $\epsilon$ whenever
$f(x,y)=0$ and at least $1-\epsilon$ whenever $f(x,y)=1$. Let
$Q_\epsilon(f)$ [$Q_\epsilon^\ast(f)$] be the smallest $c$ for which
there exists a $c$-qubit communication protocol without [respectively,
with] prior entanglement that computes $f$ with error $\epsilon$. Let
$Q(f)\df Q_{1/3}(f)$ and $Q^\ast(f)\df Q_{1/3}^\ast(f)$.

\medskip
Fix an integer $n$, and let $D\function{\{0,1,\ldots,n\}}{\{0,1\}}$ be any
Boolean predicate. Let $f_{n,D}(x,y)\df D(|x\cap y|)$, where $x,y\subseteq
[n] (\df \{1,2,\ldots,n\})$. Let $\ell_0(D)$ and $\ell_1(D)$ be given by
\refeq{l0}, \refeq{l1} (if no such $\ell$ exists, we naturally let
$\ell_\epsilon(D)\df 0$). The main result of this paper is the following
\begin{theorem} \label{main_theorem}
For every Boolean predicate $D\function{\{0,1,\ldots,n\}}{\{0,1\}}$,
$$
\Omega(\sqrt{n\ell_0(D)}+ \ell_1(D)) \leq Q^\ast(f_{n,D})\leq Q(f_{n,D})
\leq O((\sqrt{n\ell_0(D)}+ \ell_1(D))\log n).
$$
\end{theorem}
Let $DISJ_n(x,y)\df x\cap y=\emptyset$.
\begin{corollary}
$Q^\ast(DISJ_n)\geq\Omega(\sqrt n)$.
\end{corollary}

Our lower bound proof essentially uses high symmetry of the predicate
$f_{n,D}$ and, in particular, we need $x,y$ to be of the same fixed
cardinality $k$. We formulate the corresponding intermediate result in
this section since, although somewhat technical, it still might be of
independent interest.

Let $k\leq n$ and $D\function{\{0,1,\ldots,k\}}{\{0,1\}}$. Let $X=Y\df
[n]^k$ be the set of all $k$-element subsets of $[n]$ and
$f_{n,k,D}\function {X\times Y}{\{0,1\}};\ f_{n,k,D}(x,y)\df D(|x\cap y|)$
(thus, $f_{n,D}=f_{n,n,D}$).

\begin{theorem} \label{main_uniform}
Let $k\leq n/4$, $\ell\leq k/4$ and
$D\function{\{0,1,\ldots,k\}}{\{0,1\}}$ be any predicate such that
$D(\ell)\neq D(\ell-1)$. Then $Q^\ast(f_{n,k,D})\geq\Omega(\sqrt{k\ell})$.
\end{theorem}

\begin{remark} Nayak and Shi have observed (personal communication) that
our lower bound extends to a more general model in which the entanglement
need not necessarily be given in the form of shared EPR-pairs. More
specifically, in this model (considered e.g. in \cite{NaS}) the input
vector $\Input(x,y)$ has the form
\begin{equation} \label{general_input}
\Input(x,y)\df \sum_{e\in E}\lambda_e|0,x,e\rangle |0\rangle |e,y,0\rangle ,
\end{equation}
where $\set{\lambda_e}{e\in E}$ is an {\em arbitrary} unitary vector (the
case \refeq{input_string} of EPR-pairs corresponds to $\lambda_e=\frac
1{|E|^{1/2}}$). With their kind permission, we include in Section
\ref{tracenorm:sec} the adjustments to our basic proof needed for this
generalization (Remark \ref{general_entanglement}).
\end{remark}

\section{Preliminaries} \label{prel}

In this section we compile together some definitions and previously
known results needed for our proof.

\subsection{Quantum search vs. quantum communication} \label{search}

For a precise definition of a {\em quantum decision tree} see e.g.
\cite{BuW1}. Given a Boolean function $g(x_1,\ldots,x_n)$ we will denote
by $Q_{{\rm DT}}(g)$ the minimal number of queries needed to compute $g$
by a quantum decision tree with error at most 1/3 at any input
$x\in\{0,1\}^n$.

Let us denote by $f_g\function{\scr P([n])\times \scr
P([n])}{\{0,1\}}$ the predicate $f_g(x,y)\df g(x\cap y)$, where $x\cap
y$ is identified with its characteristic function. The following is
probably the deepest general fact known about quantum communication:
\begin{proposition}[\cite{BCW}] \label{simulation}
For any Boolean function $g(x_1,\ldots,x_n)$, $Q(f_g)\leq O(Q_{\rm
DT}(g)\log n)$.
\end{proposition}

\subsection{Matrix norms}
All material in this section is classical and can be found e.g. in the
excellent textbook \cite{Bha}.

After we give up Dirac's notation (in Section \ref{discrepancy:sec}), all
vectors will be represented as columns. For a complex vector $\xi$
[complex matrix $A$], let $x^\ast\df (\bar x)^\top$ [$A^\ast\df (\bar
A)^\top$, respectively] be its conjugate transpose. Let $||\xi||\df
(\xi^\ast\xi)^{1/2}$ denote the $\ell_2$-norm of $\xi$.

For a complex matrix $A$, we will denote by $||A||$ its {\em operator
norm} defined as $||A||\df\max\{||A\xi||:||\xi||\leq 1\}$. Alternatively,
$||A||=\max\{|\eta^\top A\xi|:||\eta||, ||\xi|| \leq 1\}$.

For two complex matrices $A,B$ of the same size $m\times n$ we denote by
$\langle A,B\rangle $ their entrywise scalar product, that is, $\langle
A,B\rangle \df \Tr(A^\ast B)= \sum_{i=1}^m\sum_{j=1}^n \bar a_{ij}b_{ij}$.
$\fnorm A$ denotes the {\em Frobenius norm} corresponding to this scalar
product, that is, $\fnorm A\df
\of{\sum_{i=1}^m\sum_{j=1}^n|a_{ij}|^2}^{1/2}$. We will also need the
following (somewhat more exotic) {\em trace norm} $\trnorm A$ defined as
$$
\trnorm A\df \max_B\{|\langle A,B\rangle |: ||B||\leq 1\},
$$
where $B$ runs over all (complex) matrices of the same size as $A$.

The following proposition summarizes some properties of these norms.
\begin{proposition} \label{matrix_basis}
\begin{enumerate}
\item \label{matrix_general} Let $|||\cdot|||$ be any one of the three
norms $||\cdot||$, $\fnorm\cdot$ or $\trnorm\cdot$, and $A$ be a complex
$m\times n$ matrix. Then:
\begin{enumerate}
\item $|||A^\ast|||= |||A^\perp||| = |||A|||$;
\item if $B$ is a submatrix of $A$ then $|||B|||\leq |||A|||$;
\item \label{unitary_invariant} $|||\cdot|||$ is invariant under left and
right unitary transformations, that is, for every $m\times m$ unitary
matrix $U$ and every $n\times n$ unitary matrix $V$, $|||UAV|||= |||A|||$.
\end{enumerate}
\item \label{product_norm} Let now $B$ be another complex $n\times k$ matrix,
and $AB$ stand for the ordinary matrix multiplication. Then:
\begin{enumerate}
\item \label{spectral_product} $||AB||\leq ||A||\cdot ||B||$;
\item \label{hoelder} $\trnorm{AB}\leq \fnorm A\cdot\fnorm B$. {\rm (Hoelder
inequality, see e.g. \cite[Corollary IV.2.6]{Bha})}
\end{enumerate}
\item \label{fvstwo} $||A||\leq \fnorm A\leq (\min\{m,n\})^{1/2}\cdot
||A||$.
\item For a square $n\times n$ matrix $A$, $\trnorm A\geq \sum_{i=1}^n
|a_{ii}|$.
\end{enumerate}
\end{proposition}
\begin{remark}
If $\sigma_1(A)\geq\sigma_2(A)\geq\ldots\geq \sigma_p(A),\ p=\min\{m,n\}$
are the singular values of $A$ then $||A||=\sigma_1(A)$, $\fnorm
A=\of{\sum_{t=1}^p\sigma_t^2(A)}^{1/2}$ and $\trnorm
A=\sum_{t=1}^p\sigma_t(A)$ (which, along with Proposition
\ref{matrix_basis}.\ref{unitary_invariant}) and the SVD-theorem almost
immediately implies all non-trivial parts of that proposition). We,
however, will not need this singular value characterization in our proof.
\end{remark}
\begin{remark}
The same proposition \ref{matrix_basis}.\ref{unitary_invariant}) implies
that we can unambiguously talk of the operator, Frobenius or trace norm of
an {\em operator} from one (finite-dimensional) Hilbert space to another.
\end{remark}

Two more matrix norms we will be using are the $\ell_1$-norm and
$\ell_\infty$-norm defined entrywise:
\begin{eqnarray*}
\ell_1(A) &\df& \sum_{1\leq i\leq m} \sum_{1\leq j\leq n} |a_{ij}|;\\
\ell_\infty(A) &\df& \max_{1\leq i\leq m} \max_{1\leq j\leq n} |a_{ij}|.
\end{eqnarray*}
Of course, these norms are {\em not} invariant under unitary
transformations. However, they are at least somewhat related to unitary
invariant norms via the following (obvious) observation:
$$
|\langle A,B\rangle | \leq \ell_1(A)\cdot \ell_\infty(B).
$$

\subsection{Decomposition of quantum communication protocols}

\begin{proposition}[\cite{Yao5,Kre}] \label{yao_kremer}
Let $P$ be a $c$-qubit communication protocol, and let $U_p$ be the
unitary operator in the right-hand side of \refeq{up}. Then there exist
linear operators $A_u$ on $\scr H_A$ and $B_u$ on $\scr H_B$ {\rm
(}$u\in\{0,1\}^c${\rm )} such that for every vector $a\in\scr H_A$ and
every vector $b\in\scr H_B$,
$$
U_p(|a\rangle |0\rangle |b\rangle ) =\sum_{u\in\{0,1\}^c}|A_u(a)\rangle
|u_c\rangle |B_u(b)\rangle .
$$
Moreover, $||A||, ||B||\leq 1$ for every $u\in\{0,1\}^c$.
\end{proposition}
\begin{proof}
It is only the last observation (about the operator norms) that is
(apparently) new. This, however, immediately follows from Proposition
\ref{matrix_basis}.\ref{spectral_product}) and the fact that every
operator $A_u,B_u$ is composed from unitary operators and orthogonal
projections onto the subspaces $\scr H_A\otimes|\epsilon\rangle
\otimes\scr H_B$, $\epsilon\in\{0,1\}$.
\end{proof}

\subsection{Symmetric functions and predicates} \label{symmetric}

For a Boolean predicate $D\function{\{0,1,\ldots,n\}}{\{0,1\}}$, denote by
$\widetilde\deg(D)$ the {\em approximate degree} of this predicate defined
as the minimal degree of a univariate real polynomial $f(x)$ such that
$|f(s)-D(s)|\leq 1/3$ for every $s\in \{0,1,\ldots,n\}$. Let
$g_D(x_1,\ldots,x_n)$ be the symmetric Boolean function defined as
$g_D(x)\df D\of{\sum_{i=1}^n x_i}$ (note that
$f_{n,D}(x,y)=f_{g_D}(x,y)$). \cite{NiS} observed that
$\widetilde\deg(D)= \widetilde\deg(g_D)$, where $\widetilde\deg(g)$ is
the minimal degree of a multi-variate polynomial approximating $g$ on
$\{0,1\}^n$ within error $1/3$ in the $\ell_\infty$-norm.

\begin{proposition}[\cite{Pat}] \label{paturi}
$\widetilde\deg(D)=\theta(\sqrt{n(\ell_0(D)+\ell_1(D))})$.
\end{proposition}

It was proved in \cite{BBC*} that $\Omega(\widetilde\deg(g))$ is a general
lower bound on $Q_{\rm DT}(g)$. In the opposite direction, they show that
for symmetric functions this bound is tight:
\begin{proposition}[\cite{BBC*}] \label{approx_sym}
$Q_{\rm DT}(g_D)\leq O(\widetilde\deg(D))$.
\end{proposition}

\bigskip
Assume now that $X=Y\df [n]^k$. For $0\leq s\leq k$, denote by $J_{n,k,s}$
the $0-1$ ${n\choose k}\times {n \choose k}$ matrix whose rows and columns
are indexed by $[n]^k$ and $(J_{n,k,s})|_{xy}\df \left\{ \begin{array}{l}
1\ \mbox{if}\ |x\cap y|=s\\ 0\ \mbox{otherwise}\end{array}\right.$. The
spectrum of these matrices is described by the so-called {\em Hahn
polynomials} (see e.g. \cite{Del}). The latter, being classical objects,
were re-discovered many times in different contexts; the expression that
is the most convenient for our purposes was proposed by Knuth \cite{Knu};
remarkably, it is based on a direct computation of the eigenvalues.

\begin{proposition}[\cite{Knu}] \label{eigenspaces}
Let $k\leq n/2$. The matrices $J_{n,k,s}(0\leq s\leq k)$ share the
same eigenspaces $E_0,E_1,\ldots,E_k$. The eigenvalue of $J_{n,k,s}$
corresponding to the eigenspace $E_t$ is given by
$$
\sum_{i=\max\{0,s+t-k\}}^{\min\{s,t\}} (-1)^{t-i}{t\choose i}{k-i\choose
s-i}{n-k-t+i\choose k-s-t+i}.
$$
\end{proposition}

\section{Upper bound}

In this section we show that the upper bound $Q(f_{n,D}) \leq
O((\sqrt{n\ell_0(D)}+ \ell_1(D))\log n)$ in Theorem \ref{main_theorem} is
almost immediately implied by the previously known results from Section
\ref{prel}.

Let $D\function{\{0,1,\ldots,n\}}{\{0,1\}}$ be any predicate. $D$ is
constant on the interval $[\ell_0(D),n-\ell_1(D)]$. Negating it if
necessary, we can assume that $D$ takes on value $0$ in this
interval. Then $D= D_0\lor D_1$, where $D_0^{-1}(1)\subseteq
[0,\ell_0(D)-1]$ and $D_1^{-1}(1)\subseteq
[n-\ell_1(D)+1,n]$. Also, $f_D= f_{D_0}\lor f_{D_1}$, and Alice and
Bob compute $f_{D_0}$ and $f_{D_1}$ separately.

For computing $f_{D_0}$, they apply the BCW-reduction (Proposition
\ref{simulation}) and Propositions \ref{approx_sym}, \ref{paturi}:
$$
Q(f_{D_0}) \leq O(Q_{\rm DT}(g_{D_0})\log n)\leq O(\widetilde\deg(D_0)\log
n) \leq O(\sqrt{\ell_0(D)}\log n).
$$

For computing $f_{D_1}$, Alice and Bob use the following trivial
(classical) protocol. Alice first checks whether her input $x$ has
$\leq n-\ell_1(D)$ ones or not. In the first case $f_{D_1}(x,y)=0$ and
she declares the result. Otherwise she sends her entire input to
Bob. This will take at most
$\log_2\of{\sum_{k=n-\ell_1(D)+1}^n{n\choose k}}$ bits which is
$O(\ell_1(D)\log n)$ since $\ell_1(D)\leq n/2$. Then Bob computes
$f_{D_1}(x,y)$.

\section{Lower bounds}

In this section we prove the lower bound in Theorem \ref{main_theorem}
and Theorem \ref{main_uniform}. First we show that the latter implies
the first, and this is done by a straightforward reduction.

\begin{definition} For a Boolean predicate $D$ on $\{0,1,\ldots,n\}$
and $0\leq r\leq n$, let $D-r\function{\{0,1,\ldots,n-r\}}{\{0,1\}}$ be
given by $(D-r)(s)\df D(r+s)$. Let also $D|_k$ be the restriction of $D$
onto $\{0,1,\ldots,k\}$, $k\leq n$.
\end{definition}

\begin{lemma} \label{reduction}
For every predicate $D$ on $\{0,1,\ldots,n\}$ and every integers $k,r$
satisfying $0\leq r\leq n,\ k\leq n-r$, we have $Q^\ast(f_{n,D})\geq
Q^\ast(f_{n-r,k,(D-r)|_k})$.
\end{lemma}
\begin{proof}
Alice and Bob use the optimal protocol for $f_{n,D}\function{\scr
P([n])\times \scr P([n])}{\{0,1\}}$ to compute
$f_{n-r,k,(D-r)|_k}\function{[n-r]^k\times [n-r]^k}{\{0,1\}}$. For
this they simply map their inputs $x,y\in [n-r]^k$ to the inputs
$\phi(x),\phi(y)\in \scr P([n])$ using the mapping $\phi(x)\df x\cup
\{n-r+1,\ldots,n\}$, and feed $\phi(x),\phi(y)$ into the protocol
for $f_{n,D}$.
\end{proof}

\begin{proofof}{lower bound in Theorem \ref{main_theorem} from Theorem
\ref{main_uniform}} We need to establish two separate bounds,
$Q^\ast(f_{n,D})\geq \Omega(\sqrt{n\ell_0(D)})$ and
$Q^\ast(f_{n,D})\geq \Omega(\ell_1(D))$, and both are proved via a
reduction from Lemma \ref{reduction} (with different values $r,k$ of
course). In choosing $r,k$ we must satisfy the two conditions
\begin{equation} \label{conditions}
k\leq (n-r)/4,\ (\ell-r)\leq k/4
\end{equation}
(arising from the statement of Theorem \ref{main_uniform}), where
$\ell\df\ell_0(D)$ for the first bound and $\ell\df n-\ell_1(D)$ for the
second. As long as they are satisfied, Theorem \ref{main_uniform} gives
$Q^\ast(f_{n,D})\geq \Omega(\sqrt{k(\ell-r)})$.

If $\ell\leq n/16$ (and, in particular, $\ell=\ell_0(D)$), we simply
let $r\df 0$ and $k\df n/4$. Then the bound of Theorem
\ref{main_uniform} becomes $\Omega(\sqrt{n\ell})$; i.e., exactly what
what we are proving.

If $\ell\geq n/16$, we satisfy the conditions \refeq{conditions} with
equality for which we set $r\df\frac{16\ell-n}{15}$ and $k\df \frac
4{15}(n-\ell)$. Then $\ell-r\geq\Omega(n-\ell)$, and Theorem
\ref{main_uniform} still gives us the required bound
$Q^\ast(f_{n,D})\geq\Omega(n-\ell)$.
\end{proofof}

In the rest of the paper we prove Theorem \ref{main_uniform}. The
proof splits into three fairly independent blocks.

\subsection{Approximate trace norm lower bound} \label{tracenorm:sec}

\begin{definition}
For a real matrix $M$, let $\trnorm M^\epsilon\df\min\{ \trnorm P:
\ell_\infty(M-P)\leq\epsilon \}$ be its {\em $\epsilon$-approximate trace
norm} ($P$ runs over all real matrices of the same size as $M$).
\end{definition}

\begin{definition}
For a predicate $f\function{X\times Y}{\{0,1\}}$, $M_f$ denotes its
communication 0-1 matrix $(M_f)_{xy}\df f(x,y)$.
\end{definition}

\begin{theorem} \label{tracebound}
For any predicate $f\function{X\times Y}{\{0,1\}}$, where $|X|=|Y|=N$, and
any $\epsilon>0$,
$Q_\epsilon^\ast(f)\geq\Omega(\log(\trnorm{M_f}^\epsilon/N))$.
\end{theorem}
\begin{proof}
Fix a $c$-qubit communication protocol with prior entanglement computing
$f$ with probability $\epsilon$. Let $p_{xy}$ be the acceptance
probabilities of this protocol on the input $(x,y)$; arrange them into an
$(N\times N)$ matrix $P$. Then, clearly, $\ell_\infty(M_f-P)\leq\epsilon$,
and we only have to prove that $\trnorm P\leq N\cdot\exp(O(c))$.

Apply the decomposition from Proposition \ref{yao_kremer} to the input
string \refeq{input_string}. We get:
$$
\Output(x,y) = \frac 1{|E|^{1/2}}\sum_{e\in E}\sum_{u\in \{0,1\}^c}
A_u|0,x,e\rangle |u_c\rangle B_u|e,y,0\rangle
$$
and then
\begin{eqnarray*}
p_{xy} &=& \frac 1{|E|}||\sum_{e\in E}\sum_{u\in\Pi}A_u|0,x,e\rangle
B_u|e,y,0\rangle ||^2\\ &=& \frac 1{|E|}\cdot \sum_{e,f\in E}\sum_{u,v\in
\Pi}(\langle f,x,0|A_v|A_u|0,x,e\rangle \cdot \langle
f,y,0|B_v|B_u|0,y,e\rangle ),
\end{eqnarray*}
where $\Pi\df\set{u\in \{0,1\}^c}{u_c=1}$.

Let us now define $N\times (|E|^2\times |\Pi|^2)$-matrices $A,B$ by
letting $a_{x,(efuv)}\df \langle f,x,0|A_v|A_u|0,x,e\rangle $ and
$b_{y,(efuv)}\df \langle f,y,0|B_v|B_u|0,y,e\rangle $. Then $P=\frac
1{|E|}AB^\perp$, and Proposition \ref{matrix_basis}.\ref{hoelder}) implies
\begin{equation} \label{trvsf}
\trnorm P\leq \frac 1{|E|}\cdot \fnorm A\cdot\fnorm B.
\end{equation}

For estimating $\fnorm A,\fnorm B$, we divide these matrices into $N\cdot
|\Pi|^2$ blocks, and interpret every block as an $(|E|\times |E|)$ matrix.
Namely, for any fixed $x\in X$ and $u,v\in\Pi$, let $A^{xuv}$ be the
square $(|E|\times |E|)$ matrix given by $a_{ef}^{xuv}\df a_{x,(efuv)}=
\langle f,x,0|A_v|A_u|0,x,e\rangle $. Then
\begin{equation} \label{fcomposite}
\fnorm A^2\leq N\cdot |\Pi|^2\cdot\max_{x,u,v} \fnorm{A^{xuv}}^2.
\end{equation}

For bounding $\fnorm{A^{xuv}}$ we first use Proposition
\ref{matrix_basis}.\ref{fvstwo}:
\begin{equation} \label{fvsoperator}
\fnorm{A^{xuv}} \leq |E|^{1/2}\cdot ||A^{xuv}||.
\end{equation}

Finally we claim that
\begin{equation} \label{operator_bound}
||A^{xuv}||\leq 1.
\end{equation}
Indeed, let $\eta,\xi$ be any vectors of length $|E|$ with $||\eta||,
||\xi||\leq 1$. Then we have
$$
\eta^\top A^{xuv}\xi = \langle \sum_{f\in
E}\eta_ff,x,0|A_v|A_u|0,x,\sum_{e\in E}\xi_ee\rangle
$$
and, since $||A_u||, ||A_v||\leq 1$,
\begin{eqnarray*}
||\eta^\top A^{xuv}\xi|| &\leq& ||A_u|0,x,\sum_{e\in E}\xi_ee\rangle
||\cdot ||A_v|0,x,\sum_{f\in E}\eta_ff\rangle ||\\ &\leq&
||\,|0,x,\sum_{e\in E}\xi_ee\rangle ||\cdot ||\,|0,x,\sum_{f\in
E}\eta_ff\rangle || = ||\xi||\cdot ||\eta|| \leq 1.
\end{eqnarray*}

\refeq{operator_bound} is proved. Along with \refeq{fvsoperator} and
\refeq{fcomposite} this implies $\fnorm A\leq N^{1/2}\cdot |\Pi|\cdot
|E|^{1/2}$, and the same bound holds for $\fnorm B$. Substituting them
into \refeq{trvsf}, we get $\trnorm P\leq N\cdot |\Pi|^2\leq
N\cdot\exp(O(c))$ which completes the proof of Theorem \ref{tracebound}.
\end{proof}

\begin{remark}[Nayak, Shi] \label{general_entanglement}
Theorem \ref{tracebound} (and, hence, all lower bounds following from it)
extends to the case of more general entanglement in which the input vector
is given by \refeq{general_input}. In order to see this, first note the
following generalization of the right-hand side in Proposition
\ref{matrix_basis}.\ref{fvstwo}:
\begin{equation} \label{general_frobenius}
\fnorm{LA}\leq \fnorm L\cdot ||A||
\end{equation}
(the original statement corresponds to $L=I_{\min\{m,n\}}$). If $\hat
a_{x,(efuv)}\df \lambda_ea_{x,(efuv)}$ and $\hat b_{y,(efuv)}\df
\lambda_fb_{y,(efuv)}$ then $\hat P=\hat A\hat B^\top$, where $\hat P$ is
the matrix of acceptance probabilities relative to the input vector
\refeq{general_input}, and $\trnorm{\hat P}\leq \fnorm{\hat
A}\cdot\fnorm{\hat B}$. As before, $\fnorm{\hat A}^2\leq N\cdot
|\Pi|^2\cdot\max_{x,u,v} \fnorm{\hat A^{xuv}}^2$. We, however, know that
$\hat A^{xuv} = L A^{xuv}$, where $L$ is the diagonal matrix with elements
$\set{\lambda_e}{e\in E}$. Since $\lambda$ is unitary, $\fnorm L=1$ and
\refeq{general_frobenius} implies $\fnorm{\hat A^{xuv}}\leq
||A^{xuv}||\leq 1$. The remaining calculations are the same as in the
basic proof.
\end{remark}

\subsection{Multi-dimensional discrepancy bound} \label{discrepancy:sec}

This section is central to our argument, so we begin with a brief overview
of the ordinary discrepancy bound.

Suppose that we want to get a lower bound on the approximate trace norm
(or any other approximate norm) of a matrix $M$. That is, we need to rule
out the existence of a decomposition $M=P+\Delta$, where $\trnorm P$ is
small and $\ell_\infty(\Delta)$ is small. The ordinary discrepancy method
\cite{Yao5,Kre} proceeds as follows. Assume that $M$ is a $\pm 1$-matrix,
take any probability distribution $\mu$ on its entries and form the
Hadamard product $M\circ\mu$ ($(M\circ \mu)_{ij}\df M_{ij}\mu_{ij}$). Then
$\langle M,M\circ\mu\rangle =1$ and $|\langle \Delta, M\circ\mu\rangle
|\leq \ell_1(M\circ\mu)\cdot\ell_\infty(\Delta)=\ell_\infty(\Delta)$.
Therefore, if $|\langle P, M\circ\mu\rangle |$ is small for every matrix
$P$ with small trace norm (in other words, $M\circ\mu$ has a {\em low
discrepancy} with such matrices), we are done.

The next logical step was taken by Klauck in \cite[Theorem 4]{Kla} who
observed that the ``test matrix'' need not be of the particular form
$M\circ\mu$. As long as $\mu$ is {\em any} matrix with $\ell_1(\mu)=1$ and
of low discrepancy, we are still in a good shape for {\em all} matrices
$M$ for which $|\langle M,\mu\rangle |$ is at least {\em somewhat} large.

It is well known, however, that even in this form the discrepancy method
does not work for (say) the disjointness predicate. In this paper we take
it one step further and instead of considering the linear functional
$X\mapsto \langle X,\mu\rangle $ for a {\em single} ``test matrix'' $\mu$,
we consider the multi-dimensional ``trace operator'' $X\mapsto (\langle
X,\mu_1\rangle ,\ldots, \langle X,\mu_r\rangle )$ for a {\em family} of
matrices $\mu_1,\ldots,\mu_r$ with $\ell_1(\mu_s)\leq 1$. In order to be
able to apply spectral methods, we will assume that  $\mu_1,\ldots,\mu_r$
are real symmetric commuting matrices (although it would be sufficient to
assume that they allow singular value decompositions $U\mu_1 V,\ldots,
U\mu_r V$ with common unitary matrices $U,V$).

\smallskip
\begin{definition}
An {\em $r$-dimensional discrepancy test} consists of real symmetric
matrices $\mu_1,\ldots,\mu_r$ with $\ell_1(\mu_s)\leq 1$ ($1\leq s\leq r$)
that have the same size $N\times N$ and commute with each other, along
with an orthogonal decomposition
\begin{equation} \label{decomposition}
\reals^N = E_1\oplus E_2\oplus\ldots\oplus E_k
\end{equation}
of $\reals^N$ into their shared eigenspaces $E_1,E_2,\ldots,E_k$.
\end{definition}
Note that the commutativity alone implies the existence of at least one
decomposition \refeq{decomposition}. For our application we, however, need
$k\ll N$ (that is, eigenvalues substantially repeat themselves), and for
this reason we prefer to fix the decomposition explicitly in the
definition.

Given a discrepancy test $(\mu_1,\ldots,\mu_r,E_1,\ldots,E_k)$, denote by
$\lambda_{st}$ the eigenvalue of $\mu_s$ corresponding to $E_t$. Let the
{\em trace of $E_t$} be the $r$-dimensional vector $\lambda^t$ naturally
defined as $(\lambda^t)_s\df \lambda_{st}$, and let $\Trace(\bar\mu,\bar
E)\df\set{\lambda^t\in \reals^r}{1\leq t\leq k}$ be the set of all these
vectors.

\begin{definition}
Given a set of vectors $T\subseteq \reals^r$ and $C>0$, let
$\Conv_C(T)\df\{\sum_{\lambda\in T}a_\lambda\lambda : \sum_{\lambda\in
T}|a_\lambda|\leq C\}$ be the convex hull of the segments
$\set{[-C,C]\lambda}{\lambda\in T}$. Given another vector $\xi\in\reals^r$
and $\epsilon>0$, let $\phi^\epsilon(\xi,T)\df\min\set{C}{\rho_\infty(\xi,
\Conv_C(T))\leq\epsilon}$, where $\rho_\infty$ is the distance in the
$\ell_\infty$-norm.
\end{definition}

\begin{theorem} \label{bound_on_trace}
Let $M$ be a real squared matrix, and
$(\mu_1,\ldots,\mu_r,E_1,\ldots,E_k)$ be an arbitrary $r$-dimensional test
of the same size. Let $\xi_M\in\reals^r$ be defined as $(\xi_M)_s\df
\langle M,\mu_s\rangle $. Then
$$
\trnorm M^\epsilon\geq \phi^\epsilon(\xi_M, \Trace(\bar\mu,\bar E)).
$$
\end{theorem}
\begin{proof}
Let $\trnorm M^\epsilon=C$ and $M=P+\Delta$, where $\trnorm P=C$ and
$\ell_\infty(\Delta)\leq\epsilon$. Then $\xi_M=\xi_P+\xi_\Delta$ and,
moreover, $|(\xi_\Delta)_s|= |\langle \Delta,\mu_s\rangle |\leq
\ell_1(\mu_s)\cdot \ell_\infty(\Delta)\leq\epsilon$  for every $s\in [r]$
which implies $\ell_\infty(\xi_\Delta)\leq\epsilon$. Thus, we only need to
prove that $\xi_P\in \Conv_C(\Trace(\bar\mu,\bar E))$.

Let $U$ be the orthogonal matrix corresponding to the decomposition
\refeq{decomposition}, so that all $(U^\top\mu_s U)$ are diagonal.
Consider the matrix $(U^\top PU)$, for every $t\in [k]$ let $(U^\top
PU)_t$ be its principal submatrix corresponding to the eigenspace $E_t$,
and let $a_t\df \Tr((U^\top PU)_t)$. Then $\xi_P\in
\Conv_C(\Trace(\bar\mu,\bar E))$ is implied by the following two facts:
$$
\xi_P =\sum_{t=1}^k a_t\lambda^t
$$
and
$$
\sum_{t=1}^k |a_t| \leq C.
$$
Both are proved by easy matrix manipulations (with heavy use of
Proposition \ref{matrix_basis}):
$$
(\xi_P)_s = \langle P,\mu_s\rangle  = \langle (U^\top PU), (U^\top\mu_s
U)\rangle = \sum_{t=1}^k \Tr((U^\top PU)_t)\cdot \lambda_{st}=\sum_{t=1}^k
a_t\lambda_{st}
$$
and
$$
\sum_{t=1}^k |a_t| \leq \sum_{i=1}^N |(U^\top PU)_{ii}|\leq
\trnorm{(U^\top PU)} = \trnorm P=C.
$$
\end{proof}

\subsection{Putting things together} \label{final:sec}

Now we are ready to finish the proof of Theorem \ref{main_uniform}. Fix
integers $n$ and $k\leq n/4$. Set $N\df{n\choose k}$. Let
$D\function{\{0,1,\ldots,k\}}{\{0,1\}}$ be any predicate such that
$D(\ell)\neq D(\ell-1)$ for some $\ell\leq k/4$. Applying Theorem
\ref{tracebound} (and observing that the error probability can be always
reduced from 1/3 to 1/4 with an increase in complexity by at most a
constant multiplicative factor), we get
\begin{equation} \label{step_one}
Q^\ast(f_{n,k,D})\geq\Omega(\log (\trnorm{M_{f_{n,k,D}}}^{1/4}/N)).
\end{equation}

Let now $\mu_s\df N^{-1}{k\choose s}^{-1}{n-k\choose k-s}^{-1}J_{n,k,s}$,
and let $E_0,\ldots,E_k$ be the shared eigenspaces of these matrices from
Proposition \ref{eigenspaces}. Note that $\ell_1(\mu_s)=1$ and $\langle
M_{f_{n,k,D}},\mu_s\rangle =D(s)$. Applying Theorem \ref{bound_on_trace}
with the $(k/2+1)$-dimensional test
$(\mu_0,\mu_1,\ldots,\mu_{k/2},E_0,E_1,\ldots,E_k)$, we get
\begin{equation} \label{step_two}
\trnorm{M_{f_{n,k,D}}}^{1/4} \geq \phi^{1/4}(D|_{k/2}, \Trace(\bar\mu,
\bar E)).
\end{equation}

\begin{claim}
Let $\lambda_{st}$ be the eigenvalue of the matrix $\mu_s$ corresponding
to the eigenspace $E_t$. Then:
\begin{enumerate}
\item \label{polynomiality} $\lambda_{st}= F_t(s)$, where $F_t$ is a
polynomial of degree $t$ {\rm (known, up to a normalizing factor, as Hahn
polynomial)};
\item \label{bound_on_lambda} whenever $k\leq n/4$ and
$s\leq k/2$, $|\lambda_{st}|\leq N^{-1}\cdot\exp(-\Omega(t))$.
\end{enumerate}
\end{claim}
\begin{proof}
By Proposition \ref{eigenspaces},
\begin{eqnarray*}
\lambda_{st} &=& N^{-1}{k\choose s}^{-1}{n-k\choose k-s}^{-1}\cdot
\sum_{i=\max\{0,s+t-k\}}^{\min\{s,t\}} (-1)^{t-i}{t\choose i}{k-i\choose
s-i}{n-k-t+i\choose k-s-t+i}\\ &=& N^{-1}
\sum_{i=\max\{0,s+t-k\}}^{\min\{s,t\}} (-1)^{t-i}{t\choose
i}\frac{{k-i\choose s-i}}{{k\choose s}}\frac{{n-k-t+i\choose
k-s-t+i}}{{n-k\choose k-s}}\\ &=& N^{-1} \sum_{i=0}^{t}
\left((-1)^{t-i}{t\choose i}\frac{s(s-1)\ldots(s-i+1)}{k(k-1)\ldots
(k-i+1)}\times\right.\\
&&\hspace{\mathindent}\left.\times\frac{(k-s)(k-s-1)\ldots(k-s-t+i+1)}{(n-k)(n-k-1)\ldots
(n-k-t+i+1)}\right).
\end{eqnarray*}

Part \ref{polynomiality} is already obvious from this expression. Part
\ref{bound_on_lambda} is also easy:
\begin{eqnarray*}
|\lambda_{st}| &\leq& N^{-1} \sum_{i=0}^{t}\left( {t\choose
i}\frac{s(s-1)\ldots(s-i+1)}{k(k-1)\ldots (k-i+1)}\times\right.\\
&&\hspace{\mathindent}\left.\times
\frac{(k-s)(k-s-1)\ldots(k-s-t+i+1)}{(n-k)(n-k-1)\ldots
(n-k-t+i+1)}\right)\\ &\leq& N^{-1} \sum_{i=0}^t {t\choose i}\of{\frac
sk}^i\cdot \of{\frac{k-s}{n-k}}^{t-i} = N^{-1}\cdot \of{\frac sk +
\frac{k-s}{n-k}}^t \leq N^{-1}\cdot \of{\frac 12+\frac 13}^t.
\end{eqnarray*}
\end{proof}

This claim implies that for every $t_0\leq k$, $\set{\lambda^t}{t\leq
t_0}\subseteq P(t_0)$, where $P(t_0)$ is the set of all real-valued
functions on $\{0,1,\ldots,k/2\}$ representable by (real) polynomials of
degree $\leq t_0$. Whereas $\ell_\infty(\lambda^t)\leq
N^{-1}\exp(-\Omega(t_0))$ for $t\geq t_0$. Hence,
\begin{equation} \label{inftydistance}
\forall \xi\in \Conv_C(\Trace(\bar\mu,\bar E)),\ \rho_\infty(\xi,
P(t_0))\leq N^{-1}\cdot C\cdot\exp(-\Omega(t_0)).
\end{equation}

Set now $t_0\df\widetilde\deg(D|_{k/2})-1$ and $C\df \phi^{1/4}(D|_{k/2},
\Trace(\bar\mu, \bar E))$. Note that since $\ell\leq k/4$,
\begin{equation} \label{t0}
t_0\geq\Omega(\sqrt{k\ell})
\end{equation}
by Proposition \ref{paturi}. Also, by definition of the approximate
degree, $\rho_\infty(D|_{k/2}, P(t_0))>1/3$. On the other hand, by
\refeq{inftydistance},
\begin{eqnarray*}
\rho_\infty(D|_{k/2}, P(t_0))&\leq& N^{-1}\cdot C\cdot\exp(-\Omega(t_0)) +
\rho_\infty(D|_{k/2}, \Conv_C(\Trace(\bar\mu,\bar E)))\\ &\leq&
N^{-1}\cdot C\cdot \exp(-\Omega(t_0)) + 1/4.
\end{eqnarray*}
Combining these two bounds with \refeq{t0}, we get
\begin{equation} \label{step_three}
\phi^{1/4}(D|_{k/2}, \Trace(\bar\mu,\bar E)) =C \geq N\cdot
\exp(\Omega(\sqrt{k\ell})).
\end{equation}
Theorem \ref{main_uniform} now follows from \refeq{step_one},
\refeq{step_two} and \refeq{step_three}.

\section{Acknowledgements}
I am grateful to Andris Ambainis and Avi Wigderson for several useful
discussions, to Noga Alon for pointing out the reference \cite{Knu}, to
Hartmut Klauck for pointing out an omission in the first draft of this
paper, and to Ashwin Nayak and Yaoyun Shi for their permission to include
Remark \ref{general_entanglement}.


\newcommand{\etalchar}[1]{$^{#1}$}


\end{document}